\documentstyle[11pt,aaspp4]{article}

\slugcomment{?}

\begin{document}

\title{Effects of Large-Scale Convection on $p$-mode Frequencies}

\author{M. Swisdak and E. Zweibel}
\affil{JILA, University of Colorado, Boulder, CO 80309-0440}

\begin{abstract}

We describe an approach for finding the eigenfrequencies of solar
acoustic modes ($p$ modes) in a convective envelope in the WKB limit.
This approximation restricts us to examining the effects of fluid
motions which are large compared to the mode wavelength, but allows us
to treat the three-dimensional mode as a localized ray.  The method of
adiabatic switching is then used to investigate the frequency shifts
resulting from simple perturbations to a polytropic model of the
convection zone as well as from two basic models of a convective cell.
We find that although solely depth-dependent perturbations can give
frequency shifts which are first order in the strength of the
perturbation, models of convective cells generate downward frequency
shifts which are second order in the perturbation strength.  These
results may have implications for resolving the differences between
eigenfrequencies derived from solar models and those found from
helioseismic observations.

\end{abstract}

\section{Introduction}

Comparisons between eigenfrequencies of solar acoustic modes (also
known as $p$ modes, since pressure provides the restoring force)
determined from solar models and eigenfrequencies measured via
helioseismology show that our understanding of the Sun's structure is
incomplete (\cite{gou96}).  In particular, the model frequencies show
the largest discrepancies in three distinct regions: near the solar
core, at the tachocline (the boundary between the radiative interior
and the convective envelope), and near the solar surface.  In this
work, we focus on the discrepancy near the solar surface, which is
thought to be partially a consequence of the incomplete modeling of
convective effects in solar models.

In standard solar models, the stratification of the convection zone is
determined by mixing-length theories in which one lengthscale is used
to parameterize all convective motions.  In reality, the outer
portions of the convection zone harbor turbulent convection on scales
at least as small as granules (1 Mm) and as big as supergranules
(20--30 Mm).  In addition, cells spanning the entire convection zone
(200 Mm) may also exist.  These turbulent motions affect the
frequencies and linewidths of $p$ modes, but in a manner which is not
yet fully understood.  Most authors (for example, \cite{bro84},
\cite{lav93}, \cite{gru98}) treat the lengthscales of convective
motions as small compared to the mode wavelengths.  By parameterizing
the effects of convection as a turbulent pressure, they are able to
average over the motions of individual cells, thereby simplifying the
analysis.  Our work lies at the opposite limit.  We assume that the
wavelengths of the modes are small compared to the lengthscale of the
convection.  This formulation allows us to work in the WKB limit and
treat the modes as rays following trajectories identical to those of
classical particles.

Working in the WKB limit also allows us to use the formalism of
Hamiltonian mechanics to find the $p$-mode eigenfrequencies given a
model of the convective envelope.  We introduce two related methods of
doing so, EBK (Einstein--Brillouin--Keller) quantization and adiabatic
switching.  After showing that adiabatic switching is better suited
for finding the eigenfrequencies of complex systems, we apply it to a
few simple examples of convective envelopes.

We find that for convective envelopes the fractional frequency shift
scales as the square of the ratio of the convective velocity $v$ to the
sound speed $c$.  The frequency shift averages to zero to first order in
$v/c$ and is downshifted to second order because the ray spends more
time in the region of negative Doppler shift (\cite{bro84}).  A
downshift is produced by temperature fluctuations for similar reasons.
Roughly,
\begin{equation}
-\delta\omega/\omega \sim v^2/c^2
\end{equation}
where $\delta\omega$ is the change in the eigenfrequency $\omega$.
Evaluating $c$ at the midpoint of a $p$ mode cavity gives
\begin{equation}
-\delta\omega/\omega \sim \frac{v^2}{gR_{\odot}}\ell \sim (v_1/400)^2\ell
\end{equation}
where $v_1$ is the convective velocity in km/s and $\ell$ is the mode
degree.  For typical velocities of 0.3 km/s at a depth of 10 Mm (the
cavity midpoint of a $n=1$, $\ell = 200$ $p$ mode), the fractional
frequency shift is $\approx 10^{-4}$.  Larger $\ell$ values will
produce larger shifts.  Observational determinations of $p$-mode
eigenfrequencies quote error bars of $\delta \omega / \omega \approx 5
\times 10^{-4}$, suggesting that this mechanism could produce a
detectable shift.

We develop the problem, present the theoretical justification for the
ray approximation, and provide necessary concepts from Hamiltonian
theory in \S \ref{theorsection}.  In \S \ref{methodssection} we
describe two methods for finding the eigenfrequencies of convective
envelopes, EBK quantization and adiabatic switching. Our results for
three convective structures, a vertical sound speed perturbation and
two simple models of convective cells, are presented in \S
\ref{resultssection}.  In \S \ref{dissection} we discuss our results
and outline several possible extensions of the work.

\section{Theoretical Background}\label{theorsection}

Like an organ pipe, the Sun has resonant modes of oscillation (see
\cite{gou91} for a review).  Each eigenmode is denoted, in a manner
similar to the designation of orbitals in the hydrogen atom, by its
radial order $n$, spherical harmonic degree $\ell$, and azimuthal
order $m$.  Although in general the eigenfrequency $\omega$ associated
with each mode depends on $n$, $\ell$, and $m$, the assumption of
spherical symmetry in a model creates a degeneracy so that $\omega$ is
independent of $m$. In the Sun, rotation breaks the spherical symmetry
and hence the degeneracy.

Solar acoustic modes are radially confined within a resonant cavity by
the sound speed structure and stratification of the solar interior
(the exception being modes with $\ell = 0$ which penetrate to $r =
0$). As $\ell$ increases at fixed $n$, the inner turning point of a
mode moves outward in radius.  Although there is some dependence on
$n$, the resonant cavities of $p$ modes with $\ell \gtrsim 60$ lie
completely within the convection zone and we expect that these modes
are most strongly affected by convective structures.  Since we are
primarily interested in the turbulent convection near the solar
surface, we will consider oscillations with $\ell \gtrsim 200$ which
are trapped within the outer 10\% (by radius) of the Sun.

Instead of treating the complete problem of the effects of convective
structures of arbitrary scale on the structure and frequencies of
high-$\ell$ modes, we limit the problem still further.  Following the
lead of time-distance helioseismology (\cite{duv93}), we approximate
the three-dimensional mode as a localized ray, in the process losing
any information concerning perturbations in the structure of the mode,
but allowing the ray to be described as a Hamiltonian system.  In
order to make this simplification --- equivalent to the limits of
geometrical acoustics (\cite{lan59}) and the WKB approximation --- we
must restrict ourselves to considering the effects of convective
structures which are large compared to the wavelength of the mode.

\subsection{The Modal Equation}

We begin with the linearized equations of hydrodynamics for the vector
displacement ${\mathbf \delta r}$ of a fluid parcel.  The continuity
equation is
\begin{equation}\label{continuity}
\rho' + \nabla \cdot(\rho\,{\mathbf \delta r}) = \delta\rho +
\rho \nabla\cdot({\mathbf \delta r}) = 0 
\end{equation}
and the momentum equation is
\begin{equation}
\rho\frac{\partial^2\,\mathbf{\delta r}}{\partial t^2} = - \nabla
p' + \rho \mathbf{g}' + \rho'{\mathbf g} \mbox{,}
\end{equation}
where $\delta f$ and $f'$ are the Lagrangian and Eulerian
perturbations to an equilibrium quantity $f$, and are related through
$\delta f = f' + ({\mathbf \delta r} \cdot\nabla)f$.  The
density, pressure, and gravity are denoted by $\rho$, $p$, and
$\mathbf{g}$ respectively where boldface symbols denote vector
quantities.  Throughout this work all fluid motions are assumed to be
adiabatic, hence the density and pressure perturbations are connected
by
\begin{equation}\label{adiabatic}
\delta p = \frac{\Gamma_1\,p}{\rho}\delta \rho \mbox{,}
\end{equation}
where the adiabatic exponent 
\begin{equation}\label{gamma}
\Gamma_1 \equiv \left(\frac{\partial \ln p}{\partial \ln \rho}\right)_s
\end{equation}
with the subscript $s$ indicating that the derivative is taken at
constant specific entropy.  Finally, Poisson's equation relates
${\mathbf g}'$ and $\rho'$:
\begin{equation}\label{poisson}
\nabla \cdot {\mathbf g}' = -4 \pi G \rho' \mbox{,}
\end{equation}
where $G$ is the Newtonian gravitational constant.

Furthermore, in order to simplify the calculations we:
\begin{enumerate}
\item Assume a plane-parallel geometry.  We are concerned with
high-degree modes which are trapped close to the surface and feel the
curvature of the Sun only as a small perturbation.  This assumption
allows us to treat the equilibrium gravitational acceleration as a
constant: ${\mathbf g} = g {\mathbf \hat{z}}$.
\item Assume the region of propagation takes the form of a
two-dimensional slab.  The vertical coordinate, increasing inwards
from the surface, is given by $z$ and the horizontal coordinate is
$x$.  The horizontal length of the slab, $L$, is the Sun's
circumference: $L = 2\pi R_{\odot}$.  Finally, we assume periodic
boundary conditions in the horizontal direction so that $f |_{x=0} = f
|_{x=L}$ for any variable $f$.
\item Neglect perturbations to the gravitational acceleration so that
${\mathbf g}' = 0$.  This is known as the Cowling approximation and is
justified when either $n$ or $\ell$ are large (\cite{cow41}), the latter
being the case in this work.
\end{enumerate}

It has been shown (\cite{gou93}) that, given the above assumptions,
the system of equations (\ref{continuity})--(\ref{poisson}) can be
reduced to a single equation for the variable $\Psi =
-\rho^{-1/2}\delta p = c^2\rho^{1/2} \nabla \cdot \mathbf{\delta r}$:
\begin{equation}\label{gough8.1.8}
\left( \frac{\partial^2}{\partial t^2} + \omega_c^2\right)
\frac{\partial^2\Psi}{\partial t^2} - c^2\frac{\partial^2}{\partial
t^2}\nabla^2\Psi - c^2N^2\nabla^2_h\Psi = 0 \mbox{,}
\end{equation}
where $\nabla^2_h$ is the horizontal Laplacian operator and
\begin{equation}
c^2 \equiv \frac{\Gamma_1\,p}{\rho}
\end{equation}
is the adiabatic sound speed.  The acoustic cutoff frequency
$\omega_c$ is defined by
\begin{equation}\label{truecutoff}
\omega_c^2 \equiv \frac{c^2}{4H^2}(1 + 2 {\mathbf
\hat{z}}\cdot \nabla H_{\rho})
\end{equation}
where 
\begin{equation}
H_{\rho} \equiv \left(\frac{d \ln \rho}{dz}\right)^{-1}
\end{equation}
is the density scale
height. The buoyancy, or Brunt-V\"ais\"al\"a, frequency $N$ is defined
by
\begin{equation}
N^2 \equiv g\left( \frac{1}{H_{\rho}} - \frac{g}{c^2}\right)
\end{equation}
with $N^2 < 0$ corresponding to convective instability.

A complete solution of equation (\ref{gough8.1.8}) yields both $p$ and
$g$ modes, the latter having buoyancy as their restoring force.  Since
the solar convection zone is nearly adiabatically stratified, we have
$N^2 \approx 0$ there.  Neglecting the third term in equation
(\ref{gough8.1.8}) so that only $p$ modes are allowed, we arrive at a
modified wave equation
\begin{equation}\label{gough8.1.10}
\left( \frac{\partial^2}{\partial t^2} + \omega_c^2\right) \Psi -
c^2\nabla^2\Psi =0 \mbox{.}
\end{equation}
Note that both $c$ and $\omega_c$ can be functions of position.

\subsection{The Ray Approximation}\label{raysection}

At this point, equation (\ref{gough8.1.10}) may still be solved as an
eigenvalue problem for the structure and eigenfrequencies of the $p$
modes.  In order to consider the problem in the ray limit, we must
make one further approximation.  Consider, for the moment, a simple
solar model where $\omega_c = 0$ and $c$ is a constant.  Equation
(\ref{gough8.1.10}) then reduces to the Helmholtz equation, which has
a plane-wave solution
\begin{equation}\label{plane}
\Psi = \Psi_0 e^{i( {\mathbf k \cdot r} - \omega t)} \mbox{.}
\end{equation}
The ray associated with this plane wave follows a trajectory
perpendicular to the wavefronts.

Now, consider instead a system where the wavelength of the mode is
much shorter than the lengthscale over which equilibrium quantities
(such as $c$, $p$ and $\rho$) vary.  Letting the ratio of these
lengthscales be denoted by $\Lambda^{-1}$ --- which we treat as a constant,
small parameter --- we write the solution to equation
(\ref{gough8.1.10}) as a plane wave with varying amplitude, $\Psi_0$,
and phase, $\Phi$:
\begin{equation}\label{almostplane}
\Psi = \Psi_0( {\mathbf r},t) e^{i \Lambda \Phi ({\mathbf r},t)} \mbox{.}
\end{equation}
Although we have not made any approximations thus far, the form of
equation (\ref{almostplane}) suggests that $\Psi$ is rapidly
oscillating compared to the background state, implying a solution
which may be thought of as locally planar.  These planar wavefronts
allow the motion to be approximated by a ray.

Using this {\it ansatz} and following the work of Gough (1993), we
expand equation (\ref{gough8.1.10}) and equate powers of $\Lambda$.
This process is equivalent to the WKB approximation and gives the
leading equation
\begin{equation}\label{WKB1}
\left( \frac{\partial \Phi}{\partial t}\right) ^2 - \left(
\frac{\omega_c}{\Lambda}\right) ^2 - c^2\left| \frac{\partial
\Phi}{\partial{\mathbf x}}\right| ^2 = 0 \mbox{.}
\end{equation}
Next, making the analogy between equations (\ref{plane}) and
(\ref{almostplane}), we define the local frequency $\omega$ and
wavenumber $\mathbf{k}$ as
\begin{equation}\label{analogy}
\omega \equiv -\Lambda\frac{\partial \Phi}{\partial t} \qquad
\mbox{and} \qquad {\mathbf k} \equiv \Lambda\frac{\partial
\Phi}{\partial{\mathbf x}}
\end{equation}
These identifications allow us to write equation (\ref{WKB1}) in the
form of a dispersion relation:
\begin{equation}\label{dispersion}
\omega = (c^2k^2 + \omega^2_c)^{\frac{1}{2}} \mbox{.}
\end{equation}
Furthermore, equation (\ref{analogy}) also implies equations of
motion:
\begin{equation}\label{raymotion}
\frac{d {\mathbf k}}{dt} = -\frac{\partial \omega}{\partial {\mathbf x}}
\qquad \mbox{and} \qquad\frac{d {\mathbf x}}{dt} = \frac{\partial
\omega}{\partial {\mathbf k}} \mbox{,}
\end{equation}
where $\frac{d}{dt}$ denotes the derivative with respect to time along
a ray path.

Equations (\ref{raymotion}) are Hamilton's equations for a Hamiltonian
$H = \omega$ in terms of the canonical positions $q_i = x_i$ and
momenta $p_i = k_i$. The ray's dispersion relation, $\omega({\mathbf
x},{\mathbf k},t)$, is identified as the functional form of the
Hamiltonian.  While this result may appear serendipitous, it is
well-known in quantum mechanics (\cite{gol65}) that high-frequency
solutions to the wave equation (in other words, solutions which
oscillate rapidly compared to any background variation) follow
dynamics similar to classical particles.

\subsection{Action variables and adiabatic invariants}\label{hamsection}

The identification with Hamiltonian mechanics described in the
preceding subsection allows us to take advantage of the extensive
results available in the field.  In this subsection, we review several
results from Hamiltonian theory which will be used later.  More
details can be found in a number of classical mechanics texts (for
example, \cite{gol65}, \cite{arn78}, \cite{lic83}, \cite{tab89}).

First, we present two results which follow from the structure of the
governing equations.  In any Hamiltonian system it is known that
\begin{equation}
\frac{dH}{dt} = \frac{\partial H}{\partial t} \mbox{.}
\end{equation}
Hence, for systems where the Hamiltonian is not explicitly
time-dependent, it is a constant of the motion.  Second, if the
Hamiltonian is cyclic (independent) of one of the canonical positions
$q_i$, then the corresponding canonical momentum $p_i$ is a constant
of the motion:
\begin{equation}
\frac{\partial H}{\partial q_i} = 0 \qquad \mbox{implies} \qquad
\frac{dp_i}{dt} = 0 \mbox{.}
\end{equation} 

Certain special Hamiltonians, named integrable systems, have equal
numbers of independent constants of the motion as degrees of freedom.
Such systems allow a change of variables where all of the new
canonical positions, termed angle variables, are cyclic.  The new
canonical momenta are the actions, $I_k$, and are defined as
\begin{equation}\label{action}
I_k = \frac{1}{2\pi} \oint_{C_k} {\mathbf p} \cdot d{\mathbf q} \mbox{,}
\end{equation}
for $k = 1, \ldots , n$, where $n$ is the number of degrees of freedom
of the system.  The curves $C_k$ are topologically-independent closed
paths in phase space which do not have to be ray trajectories.  Since
the angles are cyclic, each action is a constant of the motion
resulting in $n$ total constants.  Since a Hamiltonian system with $n$
degrees of freedom has a $2n$-dimensional phase space, integrability
ensures that the phase-space trajectories are confined to
$(2n-n)=n$-dimensional surfaces which have the topology of
$n$-dimensional tori.  An integrable system is placed on a particular
invariant torus by its initial conditions and cannot stray from this
torus as the system evolves in time.

A concrete example is provided by the Hamiltonian of equation
(\ref{dispersion}) in a two-dimensional geometry.  The system has two
degrees of freedom, so its phase space has four dimensions.  If we
assume that the Hamiltonian is independent of time, then the frequency
$\omega$ is a constant.  Furthermore, if we consider a system where
$c$ and $\omega_c$ are independent of one of the position coordinates,
say $x$, then the corresponding wavenumber $k_x$ is also a constant of
the motion.  The existence of these two constants ensures the
integrability of the system and restricts phase-space motions to a
two-dimensional surface, topologically equivalent to a two-dimensional
torus.

General Hamiltonian systems, however, are usually non-integrable.  The
fate of invariant tori in non-integrable systems was addressed by the
KAM (Kolmogorov-Arnol'd-Moser) theorem.  It concerns systems which are
nearly-integrable in the sense that they may be written as
\begin{equation}\label{KAM}
H = H_0({\mathbf I}) + \epsilon H_1({\mathbf I},{\mathbf \Theta}) \mbox{,}
\end{equation}
where ${\mathbf I}$ and ${\mathbf \Theta}$ are vectors of the action
and angle coordinates, respectively, and $\epsilon$ is a small
parameter.  Roughly, the KAM theorem states that for sufficiently
small values of $\epsilon$, most invariant tori are preserved. In
other words, for small perturbations from integrability, most sets of
initial conditions remain on invariant tori.  However, the destroyed
tori are distributed throughout the invariant ones in phase space and
as the perturbation increases, their density increases.  For a strong
enough perturbation, no invariant tori remain.

\section{Methods}\label{methodssection}

Having shown that within the WKB approximation we may treat $p$ modes
as a Hamiltonian system, the powerful techniques developed for such
systems become available for our use.  Instead of solving the
eigenvalue problem of equation (\ref{gough8.1.10}), we seek the
eigenfrequencies of the Hamiltonian given by equation
(\ref{dispersion}).  However, equations (\ref{raymotion}) describe the
motion of any ray; we do not know {\it a priori} which rays correspond
to eigenmodes.  We consider two methods of isolating the systems
corresponding to eigenmodes: EBK quantization and the method of
adiabatic switching.  Both of these methods have been used in chemical
physics to understand molecular spectra (\cite{pat85},
\cite{shirts87}).  Gough (1993) discussed EBK quantization of stellar
waves but did not fully implement the method.

In this paper we are interested in finding the eigenfrequencies of
non-integrable systems which can be expressed as a perturbation to an
integrable system as in equation ({\ref{KAM}).  Our base solar state,
corresponding to $H_0$ in equation (\ref{KAM}) and which we will show
below to be integrable, will be a two-dimensional, adiabatically
stratified, plane-parallel polytrope.  Polytropic systems are those
where the pressure $p$ and density $\rho$ are assumed to be related by
\begin{equation}
\frac{p}{p_0} = \left(\frac{\rho}{\rho_o}\right)^{1 + \frac{1}{\mu}} \mbox{,}
\end{equation}
where $p_0$ and $\rho_0$ are constants and $\mu$ is the polytropic
index, which for an adiabatically stratified system may be written as
$\mu = 1 + 1/\Gamma_1$ in terms of the adiabatic exponent
defined in equation (\ref{gamma}).  The outer portion of the solar
convection zone is well-approximated by a polytrope with $\mu \approx
3$.

Enforcing hydrostatic equilibrium leads to the relations
\begin{equation}\label{prhopoly}
\frac{p}{p_0} = \left( \frac{z}{z_0}\right) ^{\mu + 1} \qquad \mbox{and} \qquad
\frac{\rho}{\rho_0} = \left( \frac{z}{z_0}\right) ^{\mu}
\end{equation}
where $z_0$ is a constant.  Finally, the
adiabatic sound speed and cutoff frequency are given by
\begin{equation}\label{polysoundcutoff}
c^2 = \frac{gz}{\mu} \qquad \mbox{and} \qquad \omega^2_{c} =
\frac{\mu(\mu +2)}{4z^2}c^2
\end{equation}
where $g$ is the gravitational acceleration.

Figure \ref{normrayfig} shows a sample raypath in an adiabatically
stratified, two-dimensional polytropic envelope with periodic
horizontal boundary conditions and $\mu = 3$.  The vertical structure
of this ray corresponds to a mode with $n = 1$, $\ell = 200$ (see
below for a description of the mode numbers); however, the horizontal
structure has been altered for legibility -- it is equivalent to a
mode with $n=1, \ell=12$.  If drawn to scale, the box would be a thin
ribbon, with an aspect ratio of $\approx 300:1$.  Furthermore, the ray
would fit $\approx 80$ horizontal wavelengths into the box instead of
$\approx 5$ as displayed.  The ray is confined in the vertical
direction by two boundaries, or caustic surfaces, where $k_z = 0$.  If
we continue to trace the ray, it will eventually fill the entire
cavity.

\placefigure{normrayfig}

\subsection{EBK quantization}\label{EBKsection}

While finding the eigenvalues of a Hamiltonian is usually considered
as a quantum-mechanical problem, historically it was first approached
from the standpoint of classical mechanics.  The quantization which
led to the Bohr model of the hydrogen atom was an early attempt to
find eigenvalues through the quantization of the action (in this case,
the orbital angular momentum).  Later refinements resulted in EBK
semi-classical quantization (\cite{ein17}, \cite{bri26}, \cite{kel58})
which states that the action $I_k$ of equation (\ref{action}) is
quantized:
\begin{equation}\label{EBK}
I_k = \frac{1}{2\pi}\oint_{C_k} {\mathbf p} \cdot d{\mathbf q} =
\left( n_k + \frac{m_k}{4}\right)\mbox{.}
\end{equation}
The quantum numbers $n_k$ and $m_k$ are non-negative integers, with
$m_k$ representing the number of caustic surfaces (those surfaces
which define the envelope of possible trajectories) crossed by
$C_k$.

As an example of EBK quantization, consider the polytropic envelope
described above.  Christensen-Dalsgaard (1980) showed that the
eigenfrequencies for this system are given by
\begin{equation}\label{truefreq}
\omega^2 = \frac{2gk_x}{\mu}\left( n+ \frac{\mu}{2}\right) \qquad
\mbox{for} \qquad n = 1,2,...
\end{equation}
where $k_x = \ell/ R_{\odot}$ where $R_{\odot}$ is the solar radius
and $\ell$ is an integer.  The latter condition arises from requiring
that $\ell$ wavelengths fit around the Sun's circumference.

EBK quantization can nearly reproduce these results.  Combining the
expressions for $c^2$ and $\omega_c^2$ from equation
(\ref{polysoundcutoff}) with the dispersion relation of equation
(\ref{dispersion}), we may solve for $k_z$:
\begin{equation}\label{kz}
k_z^2 = \frac{\omega^2-\omega_c^2}{c^2} - k_x^2 = \frac{\mu
\omega^2}{gz} -\frac{\mu^2}{4 z^2}\left( 1 + \frac{2}{\mu} \right) -
k_x^2 \mbox{,}
\end{equation}

However, $\omega$ is constant since the dispersion relation does not
explicitly depend on time.  Furthermore, $k_x$ is also constant since
the unperturbed quantities are independent of $x$.  Evaluating
equation (\ref{EBK}) on a curve of constant $x$, and noticing that
this curve touches two caustics, one each at the top and bottom of the
cavity, we have
\begin{equation}\label{vertaction}
I_z = \frac{1}{\pi}\int_{z_1}^{z_2} k_z dz = \left( n -
\frac{1}{2}\right) \qquad \mbox{for} \qquad n = 1,2,\ldots
\end{equation}
where $z_1$ and $z_2$ are the upper and lower turning points of the
ray.  With $k_z$ given by equation (\ref{kz}), the integral in
equation (\ref{vertaction}) can be done analytically.  The final
result is
\begin{equation}\label{approxfreq}
\omega^2 = \frac{2g k_x}{\mu}\left[ n - \frac{1}{2} +
\frac{\mu}{2}\left(1 + \frac{2}{\mu}\right)^{\frac{1}{2}}\right]
\qquad \mbox{for} \qquad n = 1,2,\ldots
\end{equation}

If we also evaluate equation (\ref{EBK}) on a curve of constant $z$,
we see that the curve touches no caustics and hence
\begin{equation}\label{horaction}
I_x = \frac{1}{2\pi}\int_0^{2\pi R_{\odot}} k_x dx = \ell \qquad
\mbox{for} \qquad \ell = 0,1,\ldots
\end{equation}
This gives:
\begin{equation}\label{approxkx}
k_x = \frac{\ell}{R_{\odot}} \qquad \mbox{for} \qquad \ell =
0,1,\ldots
\end{equation}

Equation (\ref{approxkx}) is the correct horizontal quantization
condition.  Equations (\ref{truefreq}) and (\ref{approxfreq}) for the
vertical quantization are not the same, however.  Taking $n=1$ and
$\mu=3/2$, the relative error in the eigenfrequencies is $\approx
6\%$, with the agreement improving as either $n$ or $\mu$ increase.
The magnitude of this error is much larger than the accuracy to which
the true $p$-mode eigenfrequencies may be determined.  However, we are
investigating frequency {\it shifts}, not magnitudes, and so the small
differences between EBK frequencies and the true frequencies are
expected to have only a small effect on our results.

While EBK quantization appears to be a fruitful method for determining
the eigenfrequencies, it is not generally applicable.  Analytically
evaluating the quantization conditions of equations (\ref{vertaction})
and (\ref{horaction}) for all but extremely simple expressions for the
sound speed and cutoff frequency is difficult.  In particular, if
these parameters depend on $x$, the horizontal wavenumber $k_x$ is not
a constant of the motion and the evaluation of the integrals is
impractical.

An alternate approach implements the quantization conditions by
propagating the ray trajectories and evaluating equations
(\ref{vertaction}) and (\ref{horaction}) numerically.  By thoughtfully
choosing the paths $C_k$, the EBK equations may be reduced to
quantizing the areas of Poincar\a'e surface of sections (phase space
slices in the $p_i$-$q_i$ plane).  While Noid and Marcus (1975) used
this approach with some success to find the energy levels of an
anharmonically coupled pair of oscillators, it is not optimal for many
non-integrable systems.  As such systems move farther from
integrability, increasing numbers of invariant tori are destroyed.
The initial conditions corresponding to these tori are then free to
wander in phase space, giving the surfaces of section a characteristic
``fuzzy'' appearance, an early example of which is seen in the work of
H\a'enon and Heiles (1964).  The form of these surfaces of section
makes numerical determination of their areas very difficult.
Consequently, we seek another method for implementing the EBK
quantization rules for non-integrable systems.

\subsection{Adiabatic Switching}

Ehrenfest (1917) was among the first to propose the concept of
adiabatic switching, although his work has been superseded by modern
treatments.  In essence, his hypothesis was that if a system is
changed in a reversible, adiabatic way, then allowed motions will be
smoothly transformed to allowed motions.  In the language of \S
\ref{hamsection}, this is analogous to saying that a system will
remain on an invariant torus if it is altered in a smooth manner.  A
thorough review of the modern implementation of the method of
adiabatic switching and its applications can be found in Skodje \&
Cary (1988).

In adiabatic switching, the system is initialized in an eigenstate of
a integrable Hamiltonian for which the eigenfrequencies are known.
The system is allowed to evolve as the strength of a perturbation (not
necessarily small, although small in our case) to the original
Hamiltonian is slowly increased, eventually leaving the system under
the direction of a new, usually non-integrable, Hamiltonian for which
the eigenfrequencies are not known.  Due to the adiabatic nature of
the switch, however, the original eigenstate has slowly relaxed into
an eigenstate of the new system from which the new eigenfrequency can
then be determined.

While adiabatic switching has been used extensively, for example, to
determine the energy levels of systems with many degrees of freedom in
complicated external electromagnetic fields, it does not have a
complete mathematical justification.  For systems with one degree of
freedom, it is known that the action is an adiabatic invariant which
remains constant under slow perturbations to the Hamiltonian, where
``slow'' means ``long with respect to other characteristic timescales
of the problem''.  In the language of \S \ref{hamsection}, the initial
invariant torus can be deformed into an invariant torus of the final
state.  For systems with more than one degree of freedom, no such
theorem exists and the overall dynamics are not as well understood.
In particular, the KAM theorem tells us that some invariant tori of
the initial system will be destroyed with the introduction of
non-integrability.  What happens when the system switches through such
destroyed tori, found everywhere in phase space, is not known in
general.

Despite these considerations, the empirical justification for
adiabatic switching is quite strong.  Skodje \& Cary (1988) explore
these and other difficulties and conclude that, with some precautions,
adiabatic switching is an excellent method for implementing EBK
quantization in non-integrable systems.

The mathematical formulation of adiabatic switching is fairly
straightforward. The non-integrable Hamiltonian for which the
eigenfrequencies are desired, which in our case is a function of
position and wavenumber, is written as a sum of two terms
\begin{equation}\label{sepH}
H({\mathbf x},{\mathbf k}) = H_0 + H_1 \mbox{,}
\end{equation}
where $H_0$ is an integrable Hamiltonian for which the
eigenfrequencies are known from EBK quantization and $H_1$ is a term
which includes everything else.  Into equation (\ref{sepH}), we
introduce a time-dependent switching function $\lambda (t)$:
\begin{equation}\label{timeH}
H({\mathbf x},{\mathbf k}) = H_0 + \lambda(t)H_1
\end{equation}
which satisfies
\begin{equation}
\lambda(0) = 0 \qquad \mbox{and} \qquad \lambda(T) = 1
\end{equation}
for some time $T$ which, in order to ensure that the transition from
$H_0$ to $H$ is adiabatic, is taken to be much longer than other
timescales associated with the system (for instance, the
characteristic propagation time $t_{ray} \approx \omega^{-1}$).  In
addition, $\lambda$ is chosen such that its first few derivatives are
continuous at the endpoints.  In this work, $\lambda$ is taken to be
\begin{equation}\label{switching}
\lambda(t) = \frac{t}{T} - \frac{1}{2\pi}\sin \left( \frac{2\pi t}{T} \right)
\mbox{.}
\end{equation}
Johnson (1985) has a discussion of the effects of different switching
functions.

At $t=0$ our initial conditions are such that we satisfy the EBK
quantization conditions for the Hamiltonian $H_0$.  We then
numerically integrate Hamilton's equations of motion (\ref{raymotion})
under the influence of the Hamiltonian $H$ given in equation
(\ref{timeH}).  As $t$ increases so does $\lambda(t)$, adiabatically
switching on the non-integrable term $H_1$, and allowing the
eigenfrequency of the system to slowly adjust from its original value.
At $t=T$ we arrive at the eigenfrequency of the full Hamiltonian $H$.

\subsection{Numerical Methods}\label{numsection}

For our reference state $H_0$, we use the plane-parallel,
adiabatically stratified polytrope whose EBK eigenfrequencies are
given by equation (\ref{approxfreq}).  In order to expand our
investigation to convective motions, we include a Doppler term in the
Hamiltonian of equation (\ref{dispersion}):
\begin{equation}\label{fulldis}
\omega - {\mathbf k \cdot v} = (c^2k^2 + \omega_c^2)^{\frac{1}{2}}
\end{equation}
where ${\mathbf v}$ is the velocity field which is taken to be zero for
the reference state.

Writing the Hamiltonian in the form of equation (\ref{timeH}), we
have:
\begin{equation}
\omega = c_0\left(k^2 + \frac{\mu(\mu +2)}{4z^2}\right)^{\frac{1}{2}}
+ \lambda(t)\left[(c - c_0)\left(k^2 + \frac{\mu(\mu
+2)}{4z^2}\right)^{\frac{1}{2}} + {\mathbf k \cdot v}\right]
\end{equation}
where $c_0$ is the polytropic sound speed and the cutoff frequency of
equation (\ref{polysoundcutoff}) has been used for $\omega_c$.  The
true definition of the cutoff frequency, given in equation
(\ref{truecutoff}), also involves the density scale height $H_{\rho}$
which, in a consistent model, will be perturbed when the sound speed
is perturbed.  However, this effect should be small and we assume in
this work that $H_{\rho}$ keeps its polytropic value of $z/\mu$,
independent of any perturbations to $c$.

Although we have analytic expressions for the sound speed and
velocities of the final states in the examples we consider below, we
anticipate that in future applications we will not.  In preparation,
we discretize the propagation region and specify the sound speed and
velocity fields at each point.  The size of the discretization element
was determined through numerical experimentation.

In order to integrate equations (\ref{raymotion}), the code uses a
slightly-modified version of the DE package of Shampine \& Gordon
(1975): a variable-order, variable-timestep Adams-Moulton-Bashforth
PECE method.  This method allows the user to specify tolerances for
both the absolute and relative error.  We found that the integration
was sufficiently accurate and quick if both were set to $10^{-5}$.

We have not yet addressed how to choose the switching time $T$ of
equation (\ref{switching}).  In order to ensure adiabaticity, we must
choose $T$ to be much longer than the longest dynamical timescale
associated with the system, in our case the time to complete one
horizontal period.  At first glance it may appear that, with infinite
computer resources, increasingly better results may be gained by
letting $T \rightarrow \infty$.  This is not the case, however.
Recall that the KAM theorem guarantees that some tori will be
destroyed with any perturbation away from integrability (although most
tori remain invariant for small perturbations).  If the system
switches through one of the destroyed tori and remains there for an
extended period, the trajectory will begin to wander through phase
space in a process called Arnold diffusion (\cite{lic83}).  The rate
of Arnold diffusion depends on the strength of the perturbation, but
it is always present.  Thus, the choice of $T$ is also bounded from
above.  Following the lead of other investigators, we choose $T$ to be
$\approx 40$ horizontal periods.

Beginning from different initial conditions on the same invariant
torus should, in principle, lead to identical eigenfrequencies for the
final state.  Our numerical experiments found a small scatter,
presumably arising from the Arnold diffusion described above.  This
scatter is observed by other investigators and it has been found
empirically, although there is some mathematical justification
presented in Skodje \& Cary (1988), that better results are obtained
by implementing the procedure for several randomly distributed initial
conditions and averaging the final eigenvalues.  We typically average
20 such initial states in this work, a number which we find suitably
constrains the result.  A typical integration of one trajectory takes
$\approx 1$ hour to complete on a 194 MHz SGI Power Onyx.

Finally, we have scaled our variables in order to simplify the
numerics.  All lengths are quoted in units of $10^9$ cm and all times
in units of $10^{2.5}$ seconds.  For example, the solar surface
gravity is $g = 2.7397$ and the Sun's radius is $R_{\odot} = 69.599$.

\section{Results}\label{resultssection}
 
Having outlined the method of adiabatic switching, we now apply it to
several examples.  Although we only consider fairly elementary
problems in this work, adiabatic switching can in principle be applied
to arbitrarily complex sound speed and velocity distributions.
However, we believe that even the application to the simple systems
considered here is new.  The major physical constraint arises from
the ray approximation made in \S \ref{raysection} which allowed us to
write the system in Hamiltonian form.  Because of this approximation
the validity of the method is restricted to systems where the
perturbations from the reference state are large compared to the
lengthscale of the mode, $k^{-1}$.

\subsection{Height-Dependent Sound Speed Perturbation}\label{vertsection}

We first explore an envelope with only a vertical perturbation to the
sound speed and no velocity perturbation.  The horizontal invariance
of the final state implies that $k_x$ is a constant of the motion and
hence that the system is integrable at point during the switching.
Since all intermediate invariant tori exist, we expect adiabatic
switching to work quite well.  For a reasonably-chosen perturbation we
can check our results using EBK quantization.  In order to compare the
two methods, we:
\begin{enumerate}
\item Choose the initial conditions so that the system is in one of
the quantized states given by equation (\ref{approxfreq}).
\item Use the method of adiabatic switching to calculate the
eigenfrequency of the new system using the perturbation described
below.
\item Use equation (\ref{EBK}) to calculate numerically the
eigenfrequency of the new, perturbed system and compare with the
results from adiabatic switching.
\end{enumerate}

Consider the simple, but not particularly realistic, case where the
sound speed is close to the polytropic version given by equation
(\ref{polysoundcutoff}):
\begin{equation}\label{vertpert}
c = \left(\frac{gz}{\mu}\right)^{\frac{1}{2}}\left[ 1 + \epsilon \sin(\kappa
z)\right]\mbox{,}
\end{equation}
where $\epsilon$ is a small parameter characterizing the sound speed
variation and $\kappa$ is the perturbation wavenumber.  For the ray
approximation to be valid we must have $\kappa \ll k_z$, a requirement
which can only be met over part of the trajectory since $k_z = 0$ at
the upper and lower turning points of the cavity.  This difficulty is
merely the well-known breakdown of the WKB approximation near the
classical turning points.  We will only require that $\kappa \ll k_z$
over most of the vertical extent of the cavity.

The acoustic cutoff frequency of equation (\ref{polysoundcutoff}) will
also change to
\begin{equation}
\omega_c = \left(\frac{g(\mu +2)}{4z}\right)^{\frac{1}{2}}\left[ 1 + \epsilon
\sin(\kappa z)\right] \mbox{,}
\end{equation}
As mentioned in \S \ref{numsection}, we ignore any accompanying
perturbation to the density scale height $H_{\rho}$ arising from the sound
speed perturbation.

We compute relative frequency shifts, $\delta \omega/\omega$, via
adiabatic switching and EBK quantization and list the results in Table
\ref{tab1}.  Even for $\epsilon=0.16$, there is no appreciable
difference in the frequency shifts.  When applying adiabatic switching
to non-integrable systems, however, we expect the trajectory to
diffuse in phase space as invariant tori disappear, and thus the
computed frequency shifts will be less accurate as we increase the
perturbation strength.

\placetable{tab1}

A closer inspection of Table \ref{tab1} reveals two interesting
trends.  First, all frequency shifts are positive, and second, as the
final five rows demonstrate, the frequency shift is essentially linear
in the perturbation parameter $\epsilon$.  Following the analysis of
Gough (1993), we can explore both of these effects by expanding the
EBK quantization condition of equation (\ref{vertaction}).  We write
the shifted frequency as $\omega = \omega_0 + \delta\omega$, where
$\omega_0$ is the unperturbed frequency.  Similarly, we have from
equation (\ref{vertpert})
\begin{equation}\label{delc}
c = c_0 + \delta c = \left(\frac{gz}{\mu}\right)^{\frac{1}{2}} + \epsilon
\left(\frac{gz}{\mu}\right)^{\frac{1}{2}} \sin(\kappa z)
\end{equation}
where $c_0$ is the polytropic sound speed.

Applying the EBK quantization of equation (\ref{vertaction}) to the
final state yields
\begin{equation}\label{analypertshift}
\int_{z_1}^{z_2} k_z dz = \int_{z_1}^{z_2} \left[ \frac{(\omega_0 +
\delta \omega)^2 - \omega_c^2}{(c_0 + \delta c)^2} - k_x^2
\right]^{\frac{1}{2}} dz = \frac{\pi}{2} \mbox{,}
\end{equation}
where we have selected the mode with $n=1$, ignored any perturbations
to the cutoff frequency, and taken the integral over the original
cavity (which is assumed to remain unperturbed).  Next, assuming that
$\delta \omega$ and $\delta c$ are small parameters, we expand and
discard higher-order terms.  After rearranging, we have
\begin{equation}
\int_{z_1}^{z_2} k_{z,0}\left[ 1 + \frac{2}{c_0^2 k_{z,0}^2}\left(
\omega_0 \delta \omega - \omega_0^2 \frac{\delta c}{c_0} + \omega_c^2
\frac{\delta c}{c_0} \right) \right]^{\frac{1}{2}} dz = \frac{\pi}{2}
\mbox{,}
\end{equation}
where $k_{z,0}$ is the positive root of equation (\ref{kz}).  We
expand again and cancel the zeroth-order term, leaving (after some
algebraic simplifications)
\begin{equation}\label{expdelomega}
\frac{\delta \omega}{\omega_0} = \left[ \int_{z_1}^{z_2}\frac{\delta
c}{c_0}\frac{k_0^2}{k_{z,0}} dz\right]
\left[\int_{z_1}^{z_2}\frac{\omega_0^2}{c_0^2 k_{z,0}} dz \right]^{-1}
\mbox{.}
\end{equation}

Thus, in a rough sense, $\delta \omega$ is related to the integral of
$\delta c$ over the cavity.  In particular, since $\delta c$ as given
in equation (\ref{delc}) depends linearly on $\epsilon$, we expect the
same to be true of $\delta \omega$, an expectation confirmed by
adiabatic switching.  Furthermore, since every other term in equation
(\ref{expdelomega}) is non-negative, the sign of the frequency shift
is determined by the sign of the integral of $\delta c$ over the cavity.

Figure \ref{soundpertfig} shows the sound speed perturbation for the
parameters $\epsilon = 0.01$ and $\kappa = 0.10$.  The dashed lines
indicate the vertical extent of the unperturbed resonant cavity.  The
perturbation is positive through the cavity, leading to the upward
frequency shifts displayed in Table \ref{tab1}.

\placefigure{soundpertfig}

\subsection{Cell-like Velocity Perturbation}\label{velcellsec}

In both this subsection and the next we will explore simple models of
a convective cell.  While the structure of the cell in \S
\ref{convsection} is more realistic, its complexity hides some
important physical results.  So, as a first attempt we model a
convective cell as a velocity perturbation with no accompanying sound
speed perturbation.  Our velocity field is derived from a stream
function
\begin{equation}\label{stream1}
{\mathbf v} = {\mathbf \hat{y}} \times \nabla \left[ \epsilon
\sin\left(\frac{2 \pi x}{L} \right)\sin\left(\frac{2 \pi (z +
b)}{L}\right) \right]
\end{equation}
where $L = 2 \pi R_{\odot}$ is the length of the cavity, $\epsilon$
and $b$ are constants, and ${\mathbf \hat{y}}$ is a unit vector
perpendicular to the previously-defined $x$- and $z$-axes..  Although
this parameterization, by assuming incompressibility, violates the
conservation of mass flux, we are more concerned with the physical
insight we can derive from this model than its self-consistency.
Expanding equation (\ref{stream1}) gives the velocity field
\begin{equation}\label{velcellvelo}
v_x = \frac{\epsilon}{R_{\odot}}\sin \left(\frac{x}{R_{\odot}}\right)
\cos \left(\frac{z+b}{R_{\odot}} \right) \qquad \mbox{and} \qquad v_z = -
\frac{\epsilon}{R_{\odot}}\cos \left(\frac{x}{R_{\odot}}\right) \sin
\left(\frac{z+b}{R_{\odot}} \right) \mbox{.}
\end{equation}

Figure \ref{velcellfig} shows the streamlines for the velocity profile
given by equation (\ref{velcellvelo}) with $b = 3$.  The two dashed
lines near the top of the plot delineate the resonant cavity of the
initial eigenstate.  The WKB criterion is clearly met in the vertical
direction.  The streamlines close at lower depths, which are not shown
in order to better resolve the cavity.  Units for the $x$- and $z$-axes
are those given in \S \ref{vertsection} and the contours in each
plot are linearly spaced.

For all values of $b$ except integer multiples of $2\pi R_{\odot}$,
the vertical velocity is unphysically non-zero at $z=0$.  However, we
can see from Figure \ref{velcellfig} that by varying $b$, the effects
of propagation through different portions of the convective cell are
easily studied.  We make the offset parameter $b$ non-dimensional by
defining
\begin{equation}\label{phase}
\phi = \frac{b + z_c}{R_\odot}
\end{equation}
where $z_c$ is the mean depth of the resonant cavity.  The method of
adiabatic switching is then used to investigate the dependence of
eigenfrequencies on the parameters $\epsilon$ and $b$. Our results are
shown in Table \ref{tab2}.

\placetable{tab2}

Clearly the frequency shift depends on the strength of the
perturbation in a different manner than the system we considered in
the previous section.  Instead of a linear dependence, the shift
appears to be second-order in $\epsilon$. In addition, the frequency
shift is always downward.  We can explain both these effects through
the mathematical framework we developed earlier.

Our system contains no perturbation to the sound speed, but now the
Doppler term of equation (\ref{fulldis}) must be included when
evaluating equation (\ref{vertaction}):
\begin{equation}\label{velpertshift}
\oint k_z dz = \oint \left[ \frac{(\omega_0 + \delta \omega - {\mathbf
 k \cdot v})^2 - \omega_c^2}{c_0^2} - k_x^2 \right]^{1/2} dz = \pi
 \mbox{,}
\end{equation}
where we have explicitly kept the line integral from equation
(\ref{EBK}).  Performing the same expansion as done in \S
\ref{vertsection} while treating $\mathbf{k \cdot v}$ as a small
parameter, we arrive at
\begin{equation}\label{fracvelshift}
\frac{\delta \omega}{\omega_0} = \left[ \oint \frac{{\mathbf k \cdot
v}}{k_{z,0}c^2_0} dz\right]
\left[\oint\frac{\omega_0}{c_0^2 k_{z,0}} dz \right]^{-1}
\end{equation}

Again, we find that the frequency shift depends on the integral of the
perturbation over the cavity.  Since, as can be seen from Figure
\ref{velcellfig}, the ray does not interact with all of the convective
cell, we might expect the perturbation term not to average to zero.
However, in this case the perturbation has an almost antisymmetric
effect on rays propagating with and against the flow.  To first order
these shifts cancel, leaving $\delta \omega = 0$.  However, we expect
a non-zero second-order frequency shift, as can be shown by the
following thought experiment.  To first order, propagating with and
against the velocity field produces equal and opposite Doppler shifts
of the ray's frequency.  But, the ray takes more time to traverse
regions where the ray travels against the flow (${\mathbf k \cdot v}
<0$) than it does to cross regions where the ray and the flow are
co-directional (${\mathbf k \cdot v} > 0$).  The relative time
difference is of order $v/c$ and the value of the frequency shift is
of order $-v/c$, leading to a net downward shift of order $-v^2/c^2$.
This result is not restricted to the case considered here, but instead
applies for any perturbation where the ray spends equal times, to
first order, in upshift and downshift regions.  In particular, sound
perturbations caused by the thermal structure of a convective call
have this effect.

Instead of carrying out the expansion of equation (\ref{velpertshift})
to second order in the small parameters, we can graphically
demonstrate this result by varying the parameter $\phi$ of equation
(\ref{phase}).  By doing so, the resonant cavity will overlap
different regions of the convective cell, leading to different
frequency shifts.  Near the top and bottom of each cell, the flow is
predominantly horizontal.  When the resonant cavity overlaps this
region the horizontal component ($k_xv_x$) of equation
(\ref{fracvelshift}) will produce the largest shift.  Similarly, when
the cavity overlaps regions of strong upflows and downflows, the
vertical component ($k_zv_z$) will produce the strongest influence.

Figure \ref{secondorderpert} shows the relative frequency shift
plotted versus the phase $\phi$, for a $n=1$, $\ell=200$ mode with
$z_c = 0.87$ and $\epsilon = 0.05$.  The structure of the convective
cells in equation (\ref{stream1}) establishes that the plot is
periodic outside the plotted range of $0 \leq \phi \leq \pi$.  In this
example, $k_x$ is always larger than $|k_z|$ and so we expect the
effects from the horizontal perturbation to be the strongest.  The
points $\phi =0$ and $\phi =\pi$ correspond to configurations where
the resonant cavity lies near the vertical intersection of two cells,
and hence we would expect these points to show the largest
perturbation, which is indeed the case.  The minimum frequency shift
is found, as expected, when $\phi = \pi/2$ since for this alignment of
the cell and the resonant cavity the horizontal velocities sampled by
the ray are smallest.

\placefigure{secondorderpert}

\subsection{Convective Cell}\label{convsection}

While the cell studied in the preceding section was helpful in shaping
our intuition, it was not particularly realistic. A more consistent
model has sound speed and velocity perturbations in both the $x$ and
$z$ directions.  In this section we explore such a model and will show
that the net frequency shift is still negative and second order in the
strength of the perturbation.

Our reference state remains the adiabatically stratified,
plane-parallel polytrope described above.  We take our model for the
convective state from Shirer (1987).  It is a simple nonlinear model
of steady Rayleigh-B\a'ernard convection derived from the Boussinesq
equations in two dimensions.  We work with the solution with the
lowest mode numbers, two cells in the horizontal direction and one in
the vertical.

After converting Shirer's parameterization to the notation of this
paper, we are left with three free parameters: the depth of the
convective cell $z_T$, the thermal diffusivity $\kappa$, and its
product $\kappa \nu$ with the kinematic viscosity.  Mindful of the
constraint that the lengthscales of the fluid motions should be much
larger than the lengthscales of the oscillation, we work with the
largest possible convection cell, one the approximate depth of the
solar convection zone, $z_T = 0.3 R_{\odot}$.

Estimates of $\kappa$ and $\nu$ in the solar convection zone vary
widely over a number of orders of magnitude.  Our choice for these
terms, however, does not have a physical motivation.  Instead, we
choose these constants so that the size of the perturbations is kept
small.  A justification for this assumption comes from the observation
that the solar convection zone is nearly adiabatically stratified,
hence we expect the convective motions to be perturbations on the mean
sound speed structure.

With these choices, the sound speed becomes
\begin{equation}\label{convsound}
c^2 = \frac{gz}{\mu} + \alpha \cos\left( \frac{2\pi x}{L}\right)
\sin\left[ \pi\left( 1-\frac{z}{z_T}\right) \right] - \beta\sin\left[
2\pi\left( 1 - \frac{z}{z_T}\right) \right] \mbox{,}
\end{equation}
where $L = 2\pi R_{\odot}$ is the length of the cavity and $\alpha$
and $\beta$ are functions of the product $\kappa \nu$.  The velocity
perturbations are given by
\begin{equation}\label{velo}
v_x = \gamma \sin\left(\frac{2\pi x}{L}\right) \cos\left[ \pi\left(
1-\frac{z}{z_T}\right) \right] \qquad \mbox{and} \qquad v_z = \delta
\cos\left( \frac{2\pi x}{L}\right) \sin\left[ \pi\left(1
-\frac{z}{z_T}\right) \right] \mbox{,}
\end{equation}
where $\gamma$ and $\delta$ are functions of $\kappa$ and the product
$\kappa \nu$.  The nonlinear nature of the convection arises from the
final term in equation (\ref{convsound}) which is independent of $x$.
For a linear cell, the perturbation would oscillate in $x$.

Specifically,
\begin{equation}\label{alpha}
\alpha = \frac{g z_T}{\mu}\frac{\sqrt{8}}{\pi\,{\rm
Ra}}\frac{a^2+1}{a}\left[{\rm Ra} - {\rm Ra}_c \right]^{\frac{1}{2}}
\end{equation}
where 
\begin{equation}
a = \frac{2z_T}{L} = \frac{0.3}{\pi}\mbox{,} \qquad {\rm Ra} =
\frac{gz_T^3}{\kappa \nu \pi^4}\mbox{,} \qquad \mbox{and} \qquad {\rm
Ra}_c = \frac{(a^2+1)^3}{a^2}
\end{equation}
are the aspect ratio of the convective domain, the Rayleigh number,
and the critical Rayleigh number for the onset of convection,
respectively.  The parameter $\beta$ is defined by,
\begin{equation}
\beta = \frac{g z_T}{\mu}\frac{1}{\pi\,{\rm Ra}} \left[
{\rm Ra} - {\rm Ra}_c \right] 
\end{equation}
We also have
\begin{equation}
\gamma = \frac{\pi\sqrt{8}}{z_T (a^2+1)} \kappa \left[ {\rm Ra} - {\rm
Ra}_c \right]^{\frac{1}{2}}
\end{equation}
and
\begin{equation}\label{delta}
\delta = \frac{\pi \sqrt{8} a}{z_T (a^2+1)} \kappa \left[ {\rm Ra} -
{\rm Ra}_c \right]^{\frac{1}{2}}
\end{equation}

From the expressions in equations (\ref{alpha})--(\ref{delta}) we can
see that a critical parameter governing the sizes of the perturbation
is
\begin{equation}\label{critparam}
({\rm Ra} - {\rm Ra}_c) = \left[ \frac{gz_t^3}{\pi^4}\frac{1}{\kappa
\nu} - \frac{(a^2+1)^3}{a^2} \right] \mbox{.}
\end{equation}
In order to keep the perturbations small, we must choose the free
parameters -- $\kappa$ and $\nu$ -- such that this expression is
small, but non-zero.  For instance, choosing $\kappa \nu = 2.27$ and
$\kappa = 0.1$ gives ${\rm Ra} - {\rm Ra}_c = 9.5 \times 10^{-2}$,
$\alpha = 0.50$, $\beta = 5.1 \times 10^{-3}$, $\gamma = 0.013$ and
$\delta = 1.2 \times 10^{-3}$.  Since we are exploring the parameter
regime near the boundary of convective instability, small changes in
the parameters can have large effects on the size of the
perturbations.

Figure \ref{convcellfig} shows the streamlines for the velocity
profile given by equation (\ref{velo}) as well as the isotherms for
the temperature perturbation given by the last two terms of equation
(\ref{convsound}). The standard picture of rising warmer fluid and
sinking cooler fluid is seen to hold.  As in Figure \ref{velcellfig}
the streamlines and isotherms close at depth and are linearly spaced.

\placefigure{convcellfig}

This convective cell can be thought of as a combination of the
perturbations considered in the two previous subsections.  Comparison
of the velocity perturbations of equations (\ref{velo}) and
(\ref{velcellvelo}) show they are quite similar.  We expect the
perturbations arising from the velocity profile to be second-order by
the argument advanced earlier.  The sound speed perturbation of
equation (\ref{convsound}) is an interesting combination of a strong
perturbation which we expect to enter the frequency shift at second
order (the second term) due to its periodic horizontal structure and a
small perturbation which will enter at first order (the third term).
Our results show that it is difficult to separate the effects of these
terms, at least in the parameter regime we are considering.  The
different runs are documented in Table \ref{tab3} and show that the
frequencies are downshifted and generally are of second order in the
perturbation strength.

More interesting than the frequency shifts is the fate of the raypaths
under the influence of this perturbation.  As the most realistic model
of convection under consideration, we expect that these raypaths will
most closely approximate raypaths in the solar convection zone.
Recall from Figure \ref{normrayfig} that the raypaths and caustics for
integrable systems are quite regular.  We expect from the KAM theorem
that for small perturbations the eigenrays will remain on invariant
tori, implying that both the caustics and the raypaths will remain
similar to, although possibly deformed from, the integrable shape.  In
addition, as the strength of the perturbation increases we expect the
tori will eventually break down and the ray will begin to explore the
entire coordinate space.

Figure \ref{weirdrayfig} shows that both of these expectations are
borne out.  We have plotted sample raypaths for two different
strengths of the convective perturbation.  These raypaths were
computed by running the raypath integration routine after adiabatic
switching had completed, hence the frequency is constant in each
panel.  As in Figure \ref{normrayfig}, the horizontal structure of the
ray has been altered for legibility.  The top panel shows the raypaths
for a weak perturbation and confirms our prediction that the ray will
be confined to an invariant torus.  The bottom panel shows the result
from a stronger perturbation, the path of a ray which not confined to
an invariant torus.  If this integration was continued indefinitely,
the ray would eventually diffuse onto a near-vertical trajectory and
plunge out of the bottom of the displayed propagation region.

\section{Discussion and Future Work}\label{dissection}

We have demonstrated how to compute frequency shifts arising from
general perturbations with no special symmetries.  From our
investigations, we can draw some general conclusions concerning the
effects of convection on $p$-mode eigenfrequencies.  To first order,
the relative frequency shift is roughly proportional to the integral
of the perturbation over the resonant cavity of the eigenmode.  For
cases such as the sound speed perturbation considered in \S
\ref{vertsection} where the perturbation is always either positive or
negative, the shift is linear in the perturbation strength while the
direction of the shift depends on its sign.

Of more physical interest is the case where the perturbation is not
monotonic, but periodic.  Here, we have shown through physical and
mathematical arguments as well as numerical methods that the effect of
the perturbation vanishes to first order.  However, a second-order
perturbation does exist which has a pleasing interpretation of being
due to the extra time spent in regions of counter-propagating
material. It is intriguing to note that, as discussed in Gough {\it et
al.}~(1996), the structure of present solar models near the top of the
convection zone yields $p$-mode eigenfrequencies which are larger than
those observed by helioseismology.  This results of this paper show
shifts in the proper direction; but, we have not yet shown that the
second-order effect we discuss above is large enough to account for
these differences.

In addition, we have demonstrated that even mild perturbations can
significantly change the structure of an eigenmode's resonant cavity
despite producing only small frequency shifts.  As the perturbation
strength increases, the ray eventually begins to move chaotically and
is no longer confined in any obvious region.  It is an open question
whether this effect will persist when we investigate more realistic
convective models.  If it does, results from time-distance
helioseismology imply that the ray parameterization does work in the
Sun suggesting, perhaps, that consideration of average and not
individual raypaths would be useful.

We have shown that adiabatic switching is a viable method of
implementing the EBK quantization conditions for nonintegrable systems
and finding the eigenfrequencies of simple convective envelopes in the
WKB limit.  For integrable systems where the eigenfrequencies can be
directly computed from EBK quantization, adiabatic switching agrees
quite well.  Furthermore, adiabatic switching can find
eigenfrequencies of non-integrable systems which are impractical to
obtain through direct EBK quantization.  

Adiabatic switching is an attractive method for finding
eigenfrequencies, but care must be taken to apply it to problems for
which it is valid.  In the Sun, the true excited objects are global
modes which are not necessarily well approximated by one-dimensional
objects such as rays.  Mathematically, one may approximate a ray whose
dynamics are governed by equations (\ref{raymotion}) by superposing
eigenmodes to form a wavepacket.  But, Bogdan (1997) has shown that,
due to the finite number of excited $p$ modes, the smallest wavepacket
which can be constructed by a coherent superposition in the Sun is
$\approx 30$ Mm.  This dimension, characteristic of supergranules,
suggests a lower size limit on the structures which can be explored
using adiabatic switching.  Smaller structures are incompletely
sampled by the wavepacket, suggesting that their effects are not fully
realized.  The smallest convective structures cannot be treated at all
by WKB theory and their effects may well be modeled by turbulent
pressure (\cite{gol94}, \cite{gru98}).

Provided that they are of large enough scale, convective motions of
almost arbitrary complexity can be investigated with adiabatic
switching. A number of intriguing problems remain in this area
including the exploration of other analytic models of convective
structures such as upwellings and thermal plumes.  We are not limited
to analytic models, however.  Numerical models of convection such as
recent spherical shell simulations (\cite{gla95}, \cite{ell99}) and
convection in layers (\cite{bru98}) can also be studied with adiabatic
switching.

Along with exploring convective structures, we can also investigate
modifications to the dispersion relation of equation (\ref{fulldis}).
A fairly simple conceptual, although perhaps computationally
expensive, extension to this work would be to incorporate the third
dimension into the plane-parallel geometry.  The number of degrees of
freedom would increase by one, however the reference system would
remain integrable since another constant of the motion, $k_y$, would
also be added.  Perturbations analogous to the Doppler term of
equation (\ref{fulldis}) can also be included.  Both rotation and
magnetic fields can be expressed as additional terms in the dispersion
relation (\cite{gou93}).

Power spectra of the solar oscillations show that $p$ modes possess an
intrinsic line profile, the source and shape of which is not well
understood.  Previous works (\cite{kum94}, \cite{rox95}, \cite{ras98})
have treated the effects of intrinsic damping, noise, and source
structure.  Intriguingly, eigenfrequency linewidths are also generated
by some of the effects we have explored in this paper.  We ignored the
inherent time-dependence of solar convection in this work; recall,
however, that the frequency is a constant of the motion only for
time-independent systems.  Indeed, only by introducing $\lambda (t)$
into equation (\ref{timeH}) could adiabatic switching work.  In
addition, the Sun is almost certainly not an integrable system,
leading us to expect that a ray will diffuse in phase space as it
propagates.  Since the frequency is not a constant of the motion, its
frequency can gradually change.  We saw this effect in our work with
non-integrable systems, but averaged results from several initial
conditions to obtain a final result.  Perhaps a closer exploration of
the raw results will allow us insight into the nature of $p$-mode
linewidths.

Finally, as a byproduct of using adiabatic switching, we generate
multiple raypaths together with travel times for a given envelope.
This library of paths could prove useful in investigations of
time-distance helioseismology.

\acknowledgements This work was supported by NASA GSRP 153-0972, NSF
Grant AST-95-21779, the NASA Space Physics Theory Program, and the
SOHO/MDI Investigation.  We would also like to thank M. DeRosa,
J. Weiss, R. Skodje, J. Meiss, and D. Haber for helpful discussions.

\clearpage

\begin{deluxetable}{cccccc}
\tablecaption{Eigenfrequencies for the vertical sound speed
perturbation of \S \ref{vertsection} \label{tab1}}

\tablewidth{450pt}

\tablehead{ \colhead{$n\tablenotemark{a}$} &
\colhead{$l\tablenotemark{b}$} & \colhead{$\epsilon\tablenotemark{c}$}
& \colhead{$\kappa\tablenotemark{c}$} &
\colhead{$[\delta \omega/\omega]_{AS}\tablenotemark{d}$} &
\colhead{$[\delta \omega/\omega]_{EBK}\tablenotemark{e}$} }

\startdata 

1\tablenotemark{f} & 200 & &  & 3.5760 & 3.5760 \nl
2 & 200 &  &  & 4.2470 & 4.2470 \nl
1 & 400 &  &  & 5.0573 & 5.0573 \nl
 & & & & & \nl
1 & 200 & 0.01 & 0.10 & $8.5 \times 10^{-4}$ & $8.5 \times 10^{-4}$ \nl
2 & 200 & 0.01 & 0.05 & $5.9 \times 10^{-4}$ & $5.8 \times 10^{-4}$ \nl
1 & 400 & 0.01 & 0.05 & $2.0 \times 10^{-4}$ & $1.9 \times 10^{-4}$ \nl
1 & 200 & 0.01 & 0.05 & $4.1 \times 10^{-4}$ & $4.2 \times 10^{-4}$ \nl
1 & 200 & 0.02 & 0.05 & $8.4 \times 10^{-4}$ & $8.5 \times 10^{-4}$ \nl
1 & 200 & 0.04 & 0.05 & $1.7 \times 10^{-3}$ & $1.7 \times 10^{-3}$ \nl
1 & 200 & 0.08 & 0.05 & $3.4 \times 10^{-3}$ & $3.4 \times 10^{-3}$ \nl
1 & 200 & 0.16 & 0.05 & $6.8 \times 10^{-3}$ & $6.8 \times 10^{-3}$ \nl
\enddata

\tablenotetext{a}{Vertical mode number from equation (\ref{approxfreq})}
\tablenotetext{b}{Horizontal mode number from equation (\ref{approxkx})}
\tablenotetext{c}{Perturbation parameters from equation (\ref{vertpert})}
\tablenotetext{d}{Relative frequency shifts from adiabatic switching}
\tablenotetext{e}{Relative frequency shifts from EBK quantization}
\tablenotetext{f}{The first three rows give unperturbed frequencies
for comparison purposes}

\end{deluxetable}

\clearpage

\begin{deluxetable}{ccccc}
\tablecaption{Eigenfrequencies for the velocity perturbation of \S
\ref{velcellsec} \label{tab2}}

\tablewidth{450pt}

\tablehead{ \colhead{$n\tablenotemark{a}$} &
\colhead{$l\tablenotemark{b}$} &
\colhead{$\epsilon/R_{\odot}\tablenotemark{c}$} &
\colhead{$\phi\tablenotemark{d}$} & \colhead{$[\delta
\omega/\omega]_{AS}\tablenotemark{e}$}}

\startdata 

1 & 200 & 0.01 & 0.00 & $-9.6 \times 10^{-5}$ \nl
1 & 200 & 0.02 & 0.00 & $-3.8 \times 10^{-4}$ \nl
1 & 200 & 0.04 & 0.00 & $-1.4 \times 10^{-3}$ \nl
1 & 200 & 0.08 & 0.00 & $-5.5 \times 10^{-3}$ \nl
 & & & & \nl
1 & 200 & 0.05 & 0.00 & $-2.1 \times 10^{-3}$ \nl
1 & 200 & 0.05 & 0.39 & $-2.0 \times 10^{-3}$ \nl
1 & 200 & 0.05 & 0.79 & $-1.6 \times 10^{-3}$ \nl
1 & 200 & 0.05 & 1.18 & $-1.3 \times 10^{-3}$ \nl
1 & 200 & 0.05 & 1.57 & $-1.2 \times 10^{-3}$ \nl
1 & 200 & 0.05 & 2.09 & $-1.4 \times 10^{-3}$ \nl
1 & 200 & 0.05 & 2.36 & $-1.6 \times 10^{-3}$ \nl
1 & 200 & 0.05 & 2.62 & $-1.8 \times 10^{-3}$ \nl
1 & 200 & 0.05 & 3.14 & $-2.1 \times 10^{-3}$ \nl
\enddata

\tablenotetext{a}{Vertical mode number from equation (\ref{approxfreq})}
\tablenotetext{b}{Horizontal mode number from equation (\ref{approxkx})}
\tablenotetext{c}{Perturbation parameter from equation (\ref{velcellvelo})}
\tablenotetext{d}{Perturbation parameter from equation (\ref{phase})}
\tablenotetext{e}{Relative frequency shifts from adiabatic switching}
\end{deluxetable}

\clearpage

\begin{deluxetable}{ccccc}
\tablecaption{Eigenfrequencies for the convective cell of \S
\ref{convsection} \label{tab3}}

\tablewidth{450pt}

\tablehead{ \colhead{$n\tablenotemark{a}$} &
\colhead{$l\tablenotemark{b}$} & \colhead{$\kappa\tablenotemark{c}$}
& \colhead{$\nu\tablenotemark{d}$} &
\colhead{$[\delta \omega/\omega]_{AS}\tablenotemark{e}$}}

\startdata 

1 & 200 & 0.1 & 22.7 & $-8.6 \times 10^{-4}$ \nl
1 & 200 & 0.05 & 45.4 & $-7.5 \times 10^{-4}$ \nl
1 & 200 & 0.01 & 227 & $-7.2 \times 10^{-4}$ \nl
1 & 200 & 0.1 & 22.6 & $-4.8 \times 10^{-3}$ \nl
1 & 200 & 0.05 & 45.2 & $-4.2 \times 10^{-3}$ \nl
1 & 200 & 0.01 & 226 & $-4.0 \times 10^{-3}$ \nl
\enddata

\tablenotetext{a}{Vertical mode number from equation (\ref{approxfreq})}
\tablenotetext{b}{Horizontal mode number from equation (\ref{approxkx})}
\tablenotetext{c}{Thermal diffusivity -- see equations (\ref{alpha} -
\ref{delta})}
\tablenotetext{d}{Kinematic viscosity -- see equations (\ref{alpha} -
\ref{delta})}
\tablenotetext{e}{Relative frequency shifts from adiabatic switching}

\end{deluxetable}

\clearpage

\figcaption[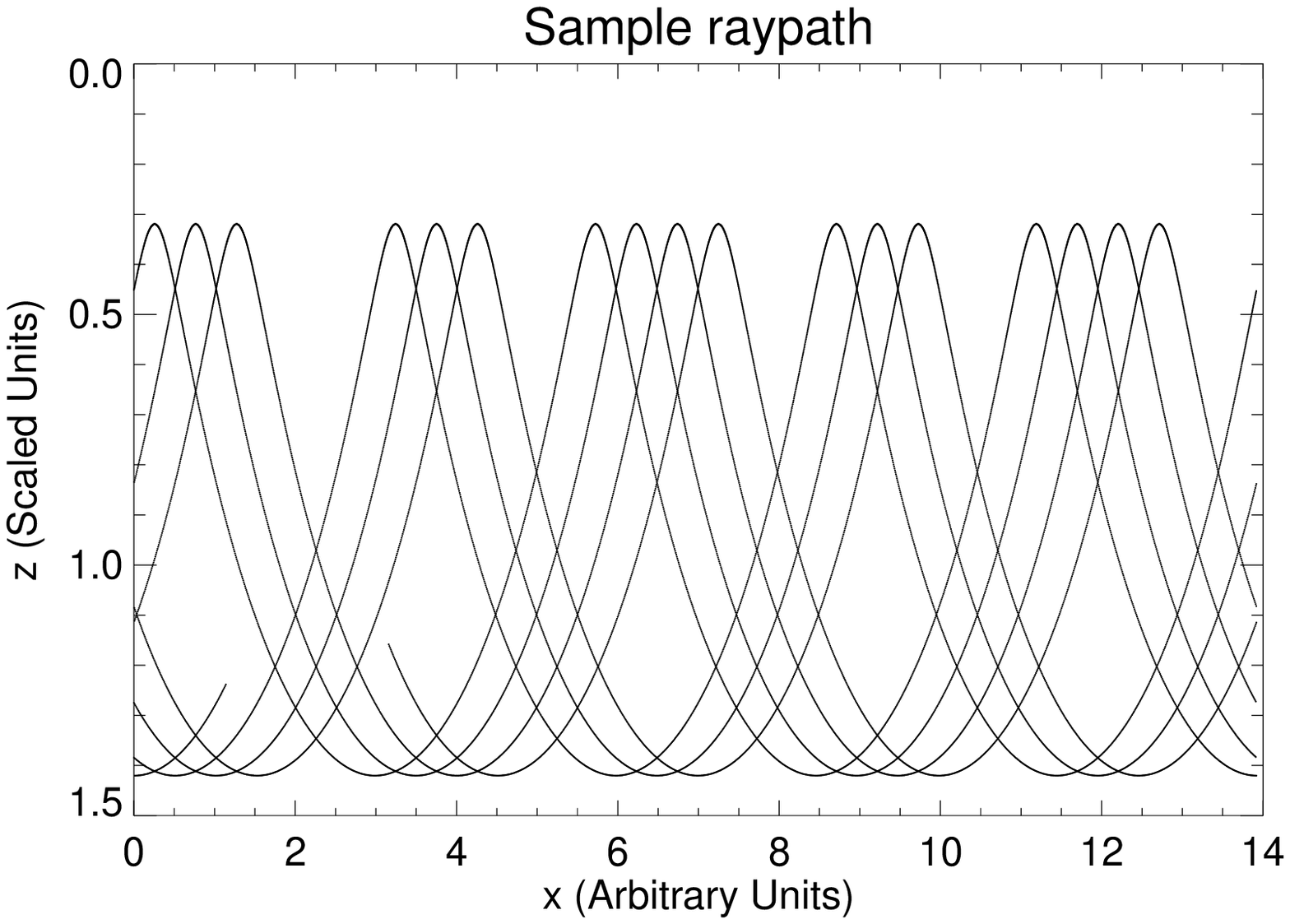]{Sample raypath in an adiabatically
stratified, plane-parallel, horizontally-periodic envelope.  The
cavity has been horizontally compressed for clarity.  Caustic surfaces
at $z = 0.32$ and $z = 1.42$ are visible. \label{normrayfig}}

\figcaption[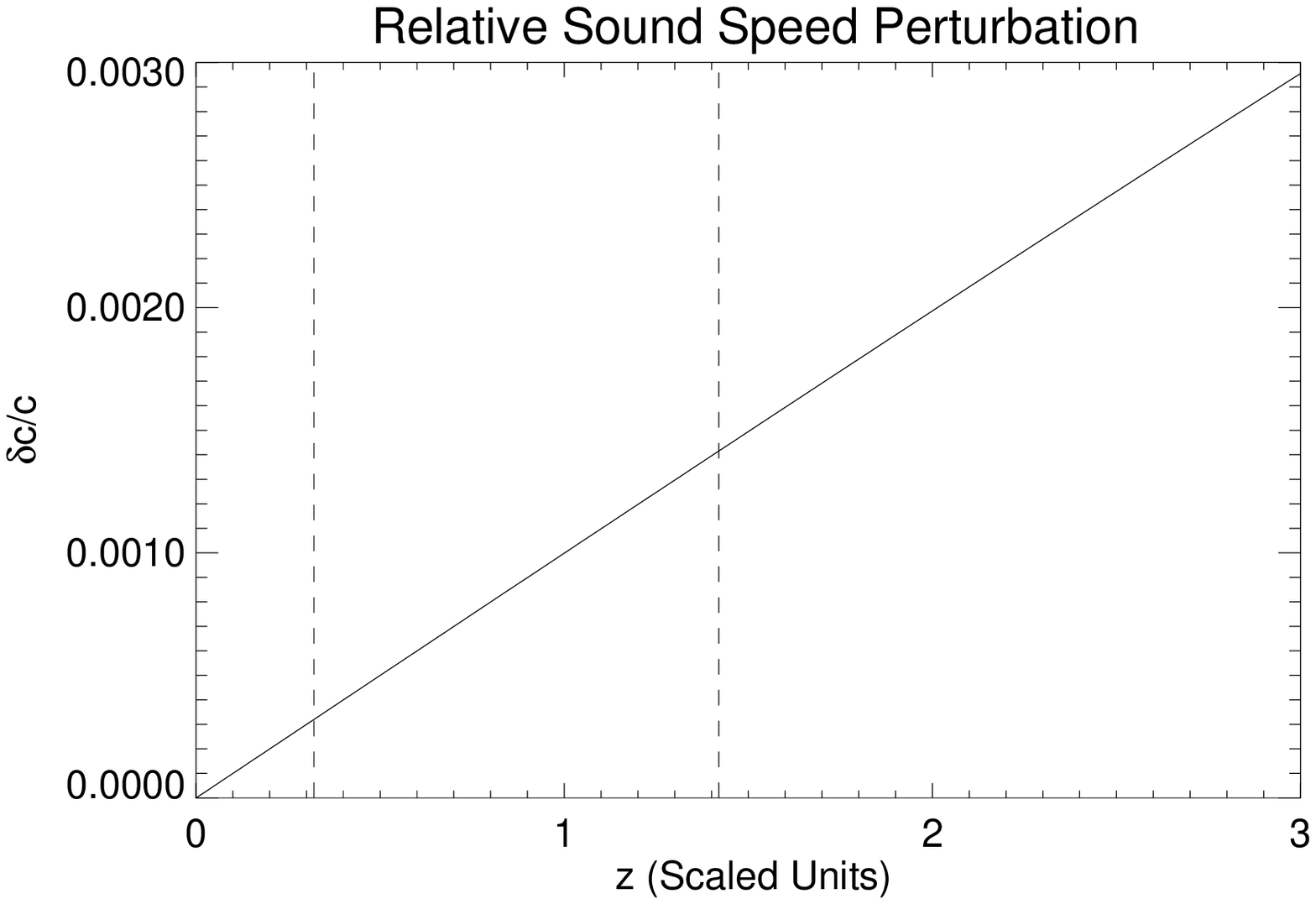]{Relative perturbation to the sound speed
described in equation (\ref{vertpert}).  The dashed lines show the
vertical extent of the ray's unperturbed cavity. \label{soundpertfig}}

\figcaption[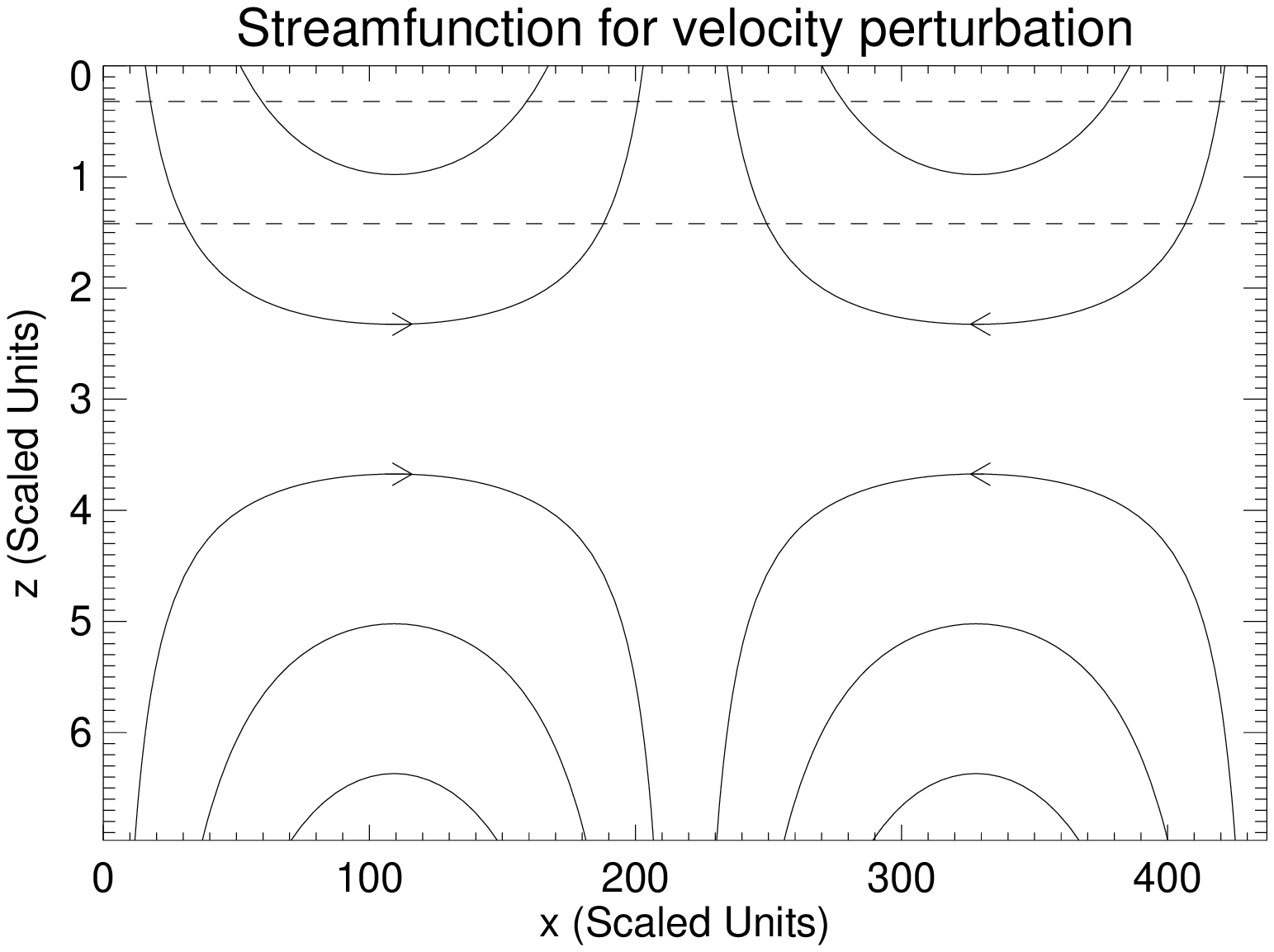]{Streamlines for the velocity perturbation
of \S \ref{velcellsec} with $b = 3$.  The streamlines connect at lower
depth and the dashed lines show the resonant cavity of the intital
ray.\label{velcellfig}}

\figcaption[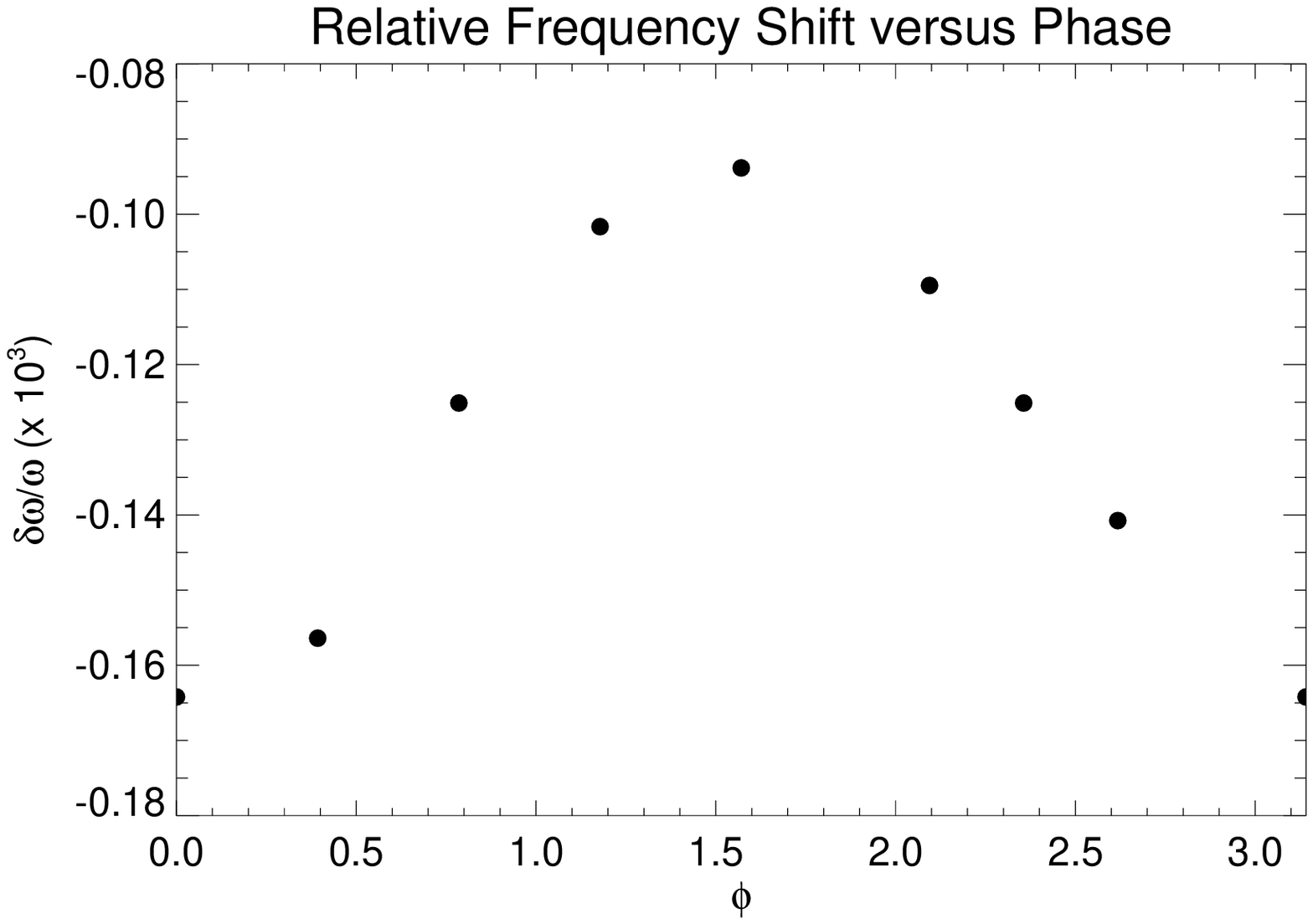]{Relative frequency shift versus the
phase of the offset between the resonant cavity and the convective
cell as defined in equation (\ref{phase}). \label{secondorderpert}}

\figcaption[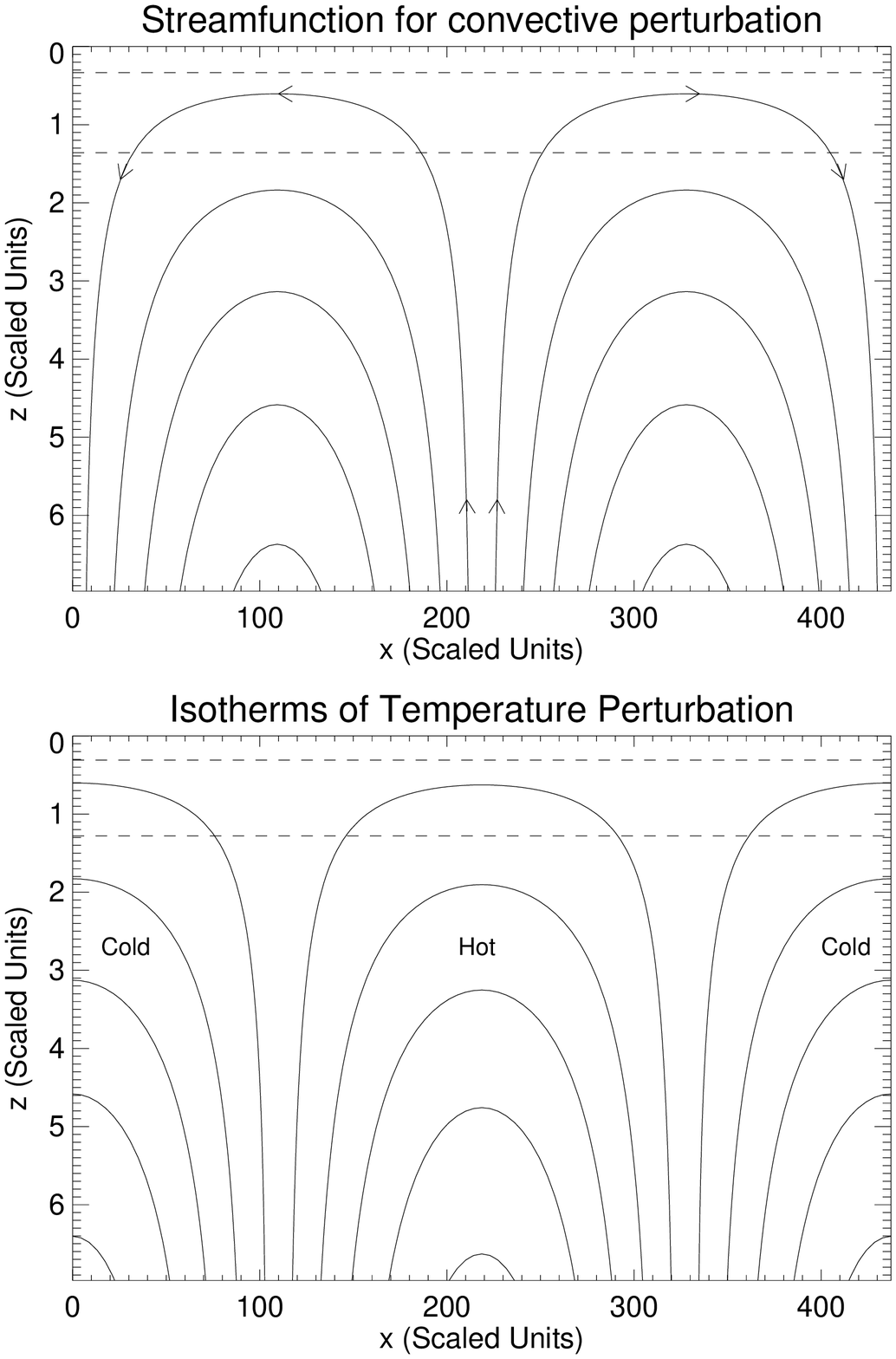]{{\it Upper panel}: Streamlines for
the convective perturbation described in \S \ref{convsection}.
The streamlines connect at lower depths.  The caustic surfaces of a
sample raypath are shown as dashed lines. {\it Lower panel}:
Temperature perturbation isotherms for the same convective
structure. \label{convcellfig}}

\figcaption[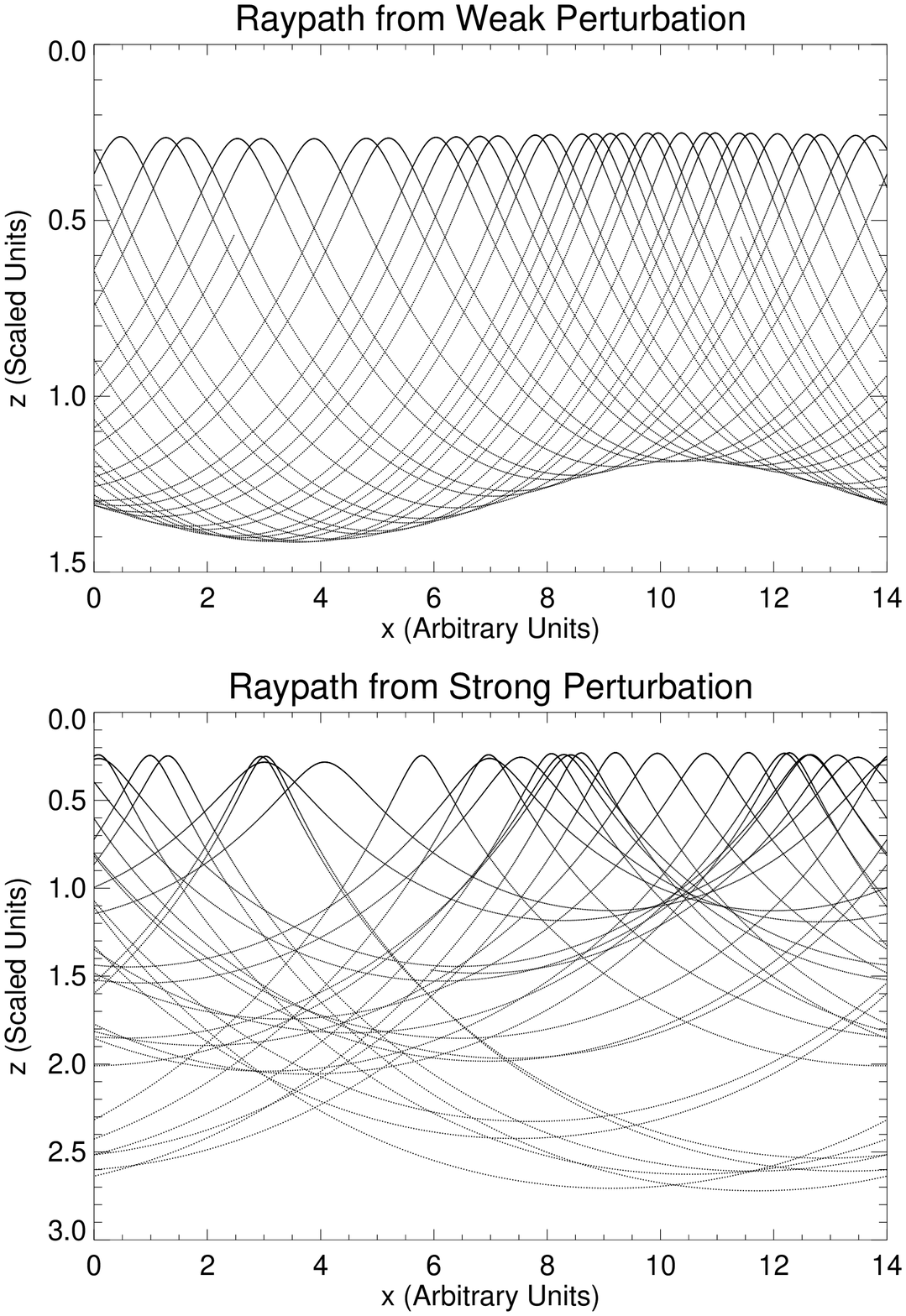]{{\it Upper panel}: Sample raypath for a weak
convective perturbation ($\kappa = 0.01$, $\nu =227$).  As in Figure
(\ref{normrayfig}), the horizontal scale has been compressed for
legibility.  The deformed caustics are clearly visible.  {\it Lower
panel}: Raypath for a stronger perturbation ($\kappa = 0.1$,
$\nu=22.6$).  The ray is no longer confined by caustics and will
eventually explore all of coordinate space. Note the difference in
vertical scales.\label{weirdrayfig}}

\clearpage

\plotone{f1.eps}

\clearpage

\plotone{f2.eps}

\clearpage

\plotone{f3.eps}

\clearpage

\plotone{f4.eps}

\clearpage

\plotone{f5.eps}

\clearpage

\plotone{f6.eps}


\clearpage


\begin{thebibliography}{}
\bibitem[Arnol'd 1978]{arn78} Arnol'd, V. I. 1978, Mathematical Methods
	of Classical Mechanics, Springer: New York
\bibitem[Bogdan 1997]{bog97} Bogdan, T. J. 1997, \apj, 477, 475
\bibitem[Brillouin 1926]{bri26} Brillouin, L. 1926, J. Phys. Radium, 7, 353
\bibitem[Brown 1984]{bro84} Brown, T. M. 1984, Science, 226, 687
\bibitem[Brummell {\it et al.} 1998]{bru98} Brummell, N. H., Hurlburt,
	N. E., \& Toomre, J. 1998, \apj, 493, 955
\bibitem[Christensen-Dalsgaard 1980]{jcd80} Christensen-Dalsgaard, J.,
	1980, \mnras, 190, 765
\bibitem[Cowling 1941]{cow41} Cowling, T. G.,
	1941, \mnras, 101, 367
\bibitem[Duvall {\it et al.} 1993]{duv93} Duvall, T. L. Jr.,
	Jefferies, S. M., Harvey, J. W., \& Pomerantz, M. A.  1993,
	\nat, 362, 430
\bibitem[Ehrenfest 1917]{ehr17} Ehrenfest, P., 1917, Phil. Mag.  33, 500
\bibitem[Einstein 1917]{ein17} Einstein, A. 1917,
	Verhandl. Deut. Physik. Ges.
\bibitem[Elliot {\it et al.} 1999]{ell99} Elliot, J. R., Miesch,
	M. S., Toomre, J., \& Glatzmaier G. A.  1998, Structure and
	Dynamics of the Interior of the Sun and Sun-Like Stars, ed. by
	S. G. Korzennik and A. Wilson, ESA Publications Division:
	Noordwijk, in press 
\bibitem[Glatzmaier \& Toomre 1995]{gla95} Glatzmaier, G. A., Toomre,
	J. 1995, Astron. Soc. Pac. Conf. Ser., 76, 200
\bibitem[Goldreich \& Murray 1994]{gol94} Goldreich, P. \& Murray,
	N. 1994, \apj, 424, 480
\bibitem[Goldstein 1965]{gol65} Goldstein, H. 1965, Classical
	Mechanics, Addison-Wesley: Reading
\bibitem[Gough 1993]{gou93} Gough, D. O.  1993, Astrophysical Fluid
	Dynamics, ed. by J.-P. Zahn \& J. Zinn-Justin, North-Holland:
	New York, 399
\bibitem[Gough {\it et al.} 1996]{gou96} Gough, D. O., Kosovichev,
	A. G., Toomre, J., Anderson, E., Antia, H. M., Basu, S.,
	Chaboyer, B., Chitre, S. M., Christensen-Dalsgaard, J.,
	Dziembowski, W. A., Eff-Darwich, A., Elliott, J. R., Giles,
	P. M., Goode, P. R., Guzik, J. A., Harvey, J. W., Hill, F.,
	Leibacher, J. W., Monteiro, M. J. P. F. G., Richard, O.,
	Sekii, T., Shibahashi, H., Takata, M., Thompson, M. J.,
	Vauclair, S., \& Vorontsov, S.V.  1996, Science, 272, 1296
\bibitem[Gough \& Toomre 1991]{gou91} Gough, D. O., \& Toomre,
	J. 1991, \araa, 29, 627
\bibitem[Gruzinov 1998]{gru98} Gruzinov, A. V. 1998, \apj, 498, 458
\bibitem[H\a'enon \& Heiles 1964]{hen64} H\a'enon, M., \& Heiles,
	C. 1964, \aj, 69, 73.
\bibitem[Johnson 1985]{joh85} Johnson, B. R. 1985, \jcp, 83,
	1204
\bibitem[Keller 1958]{kel58} Keller, J. B. 1958, Ann. Phys., 4, 180
\bibitem[Keller \& Rubinow 1960]{kel60} Keller, J. B., \&
	Rubinow, S. I. 1960, Ann. Phys., 9, 24
\bibitem[Kumar {\it et al.} 1994]{kum94} Kumar, P., Goldreich, P., and
	Kerswell, R.  1994, \apj, 427, 483
\bibitem[Landau \& Lifshitz 1959]{lan59} Landau, L. D., \& Lifshitz,
	E. M. 1959, Fluid Mechanics, Pergamon: London, 256
\bibitem[Lavely \& Ritzwoller 1993]{lav93} Lavely, E. M., \&
	Ritzwoller, M. H. 1993, \apj, 403, 810
\bibitem[Lichtenberg \& Lieberman 1983]{lic83} Lichtenberg, A. J.,
	\& Lieberman, M. A.  1983, Regular and Stochastic Motion,
	Springer-Verlag: New York
\bibitem[Noid \& Marcus 1975]{noi75} Noid, D. W., \& Marcus,
	R. A. 1975, \jcp, 62, 2119
\bibitem[Patterson 1985]{pat85} Patterson, C. W. 1985, \jcp, 83, 4618
\bibitem[Rast \& Bogdan 1998]{ras98} Rast, M. P. \& Bogdan,
	T. J. 1998, \apj, 496, 527
\bibitem[Roxburgh \& Vorontsov 1995]{rox95} Roxburgh, I. W. \&
	Vorontsov, S. V.  1995, \mnras, 272, 850
\bibitem[Shampine \& Gordon 1975]{sha75} Shampine, L. F., \&
	Gordon, M. K. 1975, Computer Solution of Ordinary Differential
	Equations, W. H. Freeman: San Francisco
\bibitem[Shirer 1987]{shi87} Shirer, H. N. 1987, Nonlinear
	Hydrodynamic Modeling: A Mathematical Introduction, ed. by
	Shirer, H. N., Springer-Verlag: New York, 22
\bibitem[Shirts {\it et al.} 1987]{shirts87} Shirts, R. B., Smiths,
	S. B., \& Patterson, C. W., \jcp, 86, 4452
\bibitem[Skodje \& Cary 1988]{sko88} Skodje, R. T., \&
	Cary, J. R. 1988, Comput. Phys. Rep., 8, 221
\bibitem[Tabor 1989]{tab89} Tabor, M. 1989, Chaos and Integrability in
	Nonlinear Dynamics: An Introduction, John Wiley \& Sons: New
	York

\end{thebibliography}
\end{document}